\documentclass[twocolumn]{ws-ijmpa}
\usepackage{float} 
\usepackage{xcolor}
\begin{document}

\title{Analytically Approximate Black Hole Solution to Higher Curvature Gravity}

\author{Seyed Naseh Sajadi}

\address{Strong Gravity Group, Department of Physics, Faculty of Science, Silpakorn University, Nakhon Pathom 73000, Thailand \\
Department of Mathematics and Computer Science,
Faculty of Science, Chulalongkorn University,  Bangkok 10330, Thailand \\
Email: naseh.sajadi@gmail.com}


\maketitle

\begin{abstract}

Higher-curvature gravity usually leads to very complicated field equations, which makes it hard to find analytical solutions. In this work, we obtain analytical charged black hole solutions in higher-curvature gravity by using black hole thermodynamics and the continued fraction expansion. We study the thermodynamics of static black holes using the first law of thermodynamics and the Smarr formula. We show that our results agree with those obtained by directly solving the field equations in Einstein gravity.

\end{abstract}


\section{Introduction}
{Einstein gravity has been remarkably successful in describing gravitational phenomena across a wide range of scales, from astrophysical systems to cosmology. Nevertheless, there are several theoretical motivations to consider extensions of general relativity. For instance, from a quantum field theory perspective, Einstein gravity is non-renormalizable and is therefore expected to be an effective low-energy description. In addition, classical solutions of general relativity generically contain spacetime singularities, such as those appearing in black holes and in the early universe. These considerations motivate the study of higher-curvature corrections, which naturally arise in effective field theory and quantum gravity frameworks. However, it is important to emphasize that, at present, there is no direct experimental evidence for a breakdown of general relativity, and possible deviations remain indirect and model-dependent—for example, in the context of inflationary dynamics or the stability properties of certain solutions. In this sense, higher-curvature gravity should be viewed as a well-motivated extension rather than a necessary replacement of Einstein’s theory \cite{Lu:2015cqa,Myers:1998gt,Capozziello:2022ygp}.}
{In addition, it is worth noting that gravity can exhibit effective repulsive behavior in certain regimes. In standard classical black holes in general relativity, gravity is usually attractive everywhere outside the horizon. However, under certain conditions—especially near the central region of singular and regular black holes, repulsive behavior can arise. These aspects have been explored in various settings in the literature\cite{Luongo:2015zaa,Luongo:2014qoa,Luongo:2010we,Luongo:2023xaw,Sajadi:2025prp} and provide further motivation for studying generalized gravitational frameworks.}
In Einstein gravity, the field equations are second order in derivatives of the metric, while, in higher-curvature gravity, the equations of motion generally contain fourth or higher-order derivatives. As a result, this increases the difficulty of finding exact black hole solutions. Therefore, since exact black hole solutions in higher-curvature gravity are generally unavailable, one must rely on numerical solutions or analytic approximation methods\cite{Lu:2015cqa,Bonanno:2019rsq,Rezzolla:2014mua}.
Although numerical methods provide accurate black hole solutions, but they are subject to several constraints. They typically probe only limited regions of parameter space and may miss entire branches of solutions. This has motivated the development of approximation analytic methods, such as continued-fraction expansions that match near-horizon and asymptotic behavior, which make the dependence on coupling constants explicit. Such approximations not only capture the essential physics of higher-derivative corrections but also enable direct comparison with numerical results, providing both theoretical insight and practical computational efficiency\cite{Kokkotas:2017zwt,Konoplya:2019ppy,Sajadi:2025nkm,Sajadi:2020axg,Sajadi:2025tnn,Sajadi:2022ybs,Ponglertsakul:2026whw,Sajadi:2025svc,Sajadi:2025kah}.

The paper is organized as follows: in Section \ref{sec2}, we present the basic formalism for a generic theory of gravity. The special cases for Einstein gravity is studied in Subsection \ref{sec2.1}. In Section \ref{con}, we summarize our findings.

\section{Theoretical Framework}\label{sec2}

The general action in four dimensions is given by
\begin{equation}
\mathcal{S}=\frac{1}{k^{2}}\int d^{4}x\,\sqrt{-g}\,\left(\mathcal{L}+\mathcal{L}_{m}\right),
\end{equation}
where $k$ is the gravitational coupling constant. We consider the gravitational part of the Lagrangian to be
\begin{equation}
\mathcal{L}=\mathcal{R}-2\Lambda+\alpha\,\mathcal{L}_{1},
\end{equation}
where $\mathcal{R}$ is the Ricci scalar, $\Lambda$ is the cosmological constant, $\alpha$ is the higher-curvature coupling constant, $\mathcal{L}_{1}$ denotes the higher-curvature correction to the gravitational Lagrangian, and $\mathcal{L}_{m}$ represents the Lagrangian density of the matter fields. {In the present work, the matter Lagrangian density is not restricted to vanish and can correspond to a general matter sector, including scalar fields. The gravitational Lagrangian $\mathcal{L}$ can also correspond to non-curvature-based theories of gravity, such as $f(Q)$ and $f(T)$ gravity\cite{Cai:2015emx,Heisenberg:2023lru}.}

 By varying the action with respect to the metric tensor, the corresponding equations of motion are obtained as ($k=1$)
\begin{align}
\mathcal{E}_{ab} = \mathcal{T}_{ab},
\end{align}
where
\begin{align}
\mathcal{E}_{ab}
&= \frac{1}{\sqrt{-g}}\,\frac{\delta \left( \sqrt{-g}\,\mathcal{L} \right)}{\delta g^{ab}}
= \mathcal{P}_{acde}\,\mathcal{R}_{b}{}^{cde}
- \frac{1}{2}\,g_{ab}\,\mathcal{L}
- 2\,\nabla^{c}\nabla^{d}\mathcal{P}_{acdb},
\label{LHS}
\\
\mathcal{T}_{ab}
&= -\frac{2}{\sqrt{-g}}\,\frac{\delta \left( \sqrt{-g}\,\mathcal{L}_{m} \right)}{\delta g^{ab}},\qquad \mathcal{P}^{abcd} \equiv \frac{\partial \mathcal{L}}{\partial \mathcal{R}_{abcd}} ,
\end{align}
and $\mathcal{T}_{ab}$ is the energy--momentum tensor, which is determined once the Lagrangian of matter $\mathcal{L}_{m}$ is specified. We consider a static, spherically symmetric black hole spacetime described by the line element
\begin{equation}
ds^{2}=-f(r)\,dt^{2}+\frac{dr^{2}}{h(r)}+r^{2}d\theta^{2}+r^{2}\sin^{2}\theta\,d\phi^{2},
\end{equation}
where $f(r)$ and $h(r)$ are two independent metric functions. Substituting this ansatz into the field equations~\eqref{LHS} leads to highly complicated higher-order differential equations, even in vacuum, making exact analytical solutions difficult to obtain. We therefore seek approximate solutions.
Assuming the black hole solution with event horizon located at $r=r_{+}$, we expand the metric functions in a Taylor series about the horizon,
\begin{align}
f(r) &= \sum_{i=1}^{\infty} f_{i}\,(r-r_{+})^{i}
= f_{1}(r-r_{+})+f_{2}(r-r_{+})^{2}+f_{3}(r-r_{+})^{3}+\cdots, \\
h(r) &= \sum_{i=1}^{\infty} h_{i}\,(r-r_{+})^{i}
= h_{1}(r-r_{+})+h_{2}(r-r_{+})^{2}+h_{3}(r-r_{+})^{3}+\cdots .
\end{align}
The coefficients $h_i$ and $f_i$ can be determined by inserting these expansions into the field equations.
Moreover, at large $r$, we assume
\begin{align}
h(r) = H_0 + \frac{H_1}{r} + \frac{H_2}{r^2} + ...,\;\;\;\;\;
f(r) = F_0 + \frac{F_1}{r} + \frac{F_2}{r^2} + ...,
\end{align}
where $H_i$ and $F_i$ are undetermined constants. 
We wish to obtain an approximate analytical solution that is valid near the horizon
and at large $r$. To this end, we employ a continued fraction expansion by letting \cite{Rezzolla:2014mua,Kokkotas:2017zwt,Konoplya:2019ppy,Zinhailo:2018ska}
\begin{equation}\label{eqq11}
h(r)=xA(x),\hspace{0.5cm}\dfrac{h(r)}{f(r)}=B^{2}(x),
\end{equation}
with
\begin{align}
A(x) &=1-\epsilon(1-x)+(a_{0}-\epsilon)(1-x)^{2}+\tilde{A}(x)(1-x)^{3},
\label{Ax}
\\
B(x) &=1+b_{0}(1-x)+\tilde{B}(x)(1-x)^{2},
\label{Bx}
\end{align} 
where
\begin{equation}
x = 1- \frac{r_+}{r}, \qquad 
\tilde{A}(x)=\dfrac{a_{1}}{1+\dfrac{a_{2}x}{1+\dfrac{a_{3}x}{1+\dfrac{a_{4}x}{1+...}}}},
\qquad 
\tilde{B}(x)=\dfrac{b_{1}}{1+\dfrac{b_{2}x}{1+\dfrac{b_{3}x}{1+\dfrac{b_{4}x}{1+...}}}}.
\label{cfrac}
\end{equation}
Unlike other expansions, which are valid only in a finite radius of convergence, continued fractions can remain accurate from the event horizon to spatial infinity. Another advantage of this method is its theory independence. i.e, any gravitational theories can be studied by only changing the coefficients, while the general structure of the expansion is fixed. {Although the original continued-fraction formalism was developed for asymptotically flat spacetimes, it can be generalized to asymptotic AdS and scalar-hairy black holes by separating the known asymptotic behavior and applying the continued-fraction expansion only to the residual regular functions \cite{Konoplya:2022kld}.} Despite its strengths, however, this method has some limitations; the accuracy of the solution depends on the number of coefficients considered in the expansion. Truncating the expansion introduces an error into the solution. The second limitation is that this method only describes the outside of the black hole with multiple horizons. 
Note that, here, we truncate the continued fraction at order four. 
By expanding \eqref{eqq11} near the asymptotic region ($ x\to 1 $), we obtain the following relations,  
\begin{equation}
\epsilon=-\dfrac{H_{1}}{r_+}-1,  \qquad b_{0}=\dfrac{F_1-H_1}{2r_{+}},  \qquad a_{0}=\dfrac{H_{2}}{r_{+}^{2}},
\end{equation}
and by expanding near the horizon $x\to 0$
\begin{align}
 a_1&=-1-a_{0}+2\epsilon+r_{+}h_1,\quad  a_{2}=-\dfrac{1}{ a_1} \left[4a_1-5\epsilon+1+3 a_{0}+ h_{2}r_{+}^2\right]
\nonumber \\
b_1 &= -1+\sqrt{\dfrac{h_1}{f_1}},\quad b_{2}=\dfrac{(-4f_{1}+f_{2}r_{+})b_{1}^2+2(-2f_{1}+f_{2}r_{+})b_{1}+r_{+}(f_{2}-h_{2})}{2f_{1}b_{1}(1+b_{1})},\nonumber
\end{align}
for the lowest order expansion coefficients.
Generally, by substituting these back into the expansions, the metric functions can be determined. However, the coefficients still contain two unknown near-horizon parameters, $f_{1}$ and $h_{1}$, and two asymptotic parameters, $F_{1}$ and $H_{1}$. The next step is therefore to fix these coefficients. To do so, we employ black hole thermodynamics, in particular the first law of thermodynamics and the Smarr relation.  
From the near-horizon expansion, the Hawking temperature is obtained as
\begin{equation}\label{eqTe}
T=\frac{\kappa}{2\pi}=\frac{\sqrt{f_{1}h_{1}}}{4\pi},
\qquad
\kappa^{2}=-\frac{1}{2}\nabla_{\alpha}\xi_{\beta}\nabla^{\alpha}\xi^{\beta}\bigg|_{r=r_{+}},
\end{equation}
where $\xi^{\mu}$ is the timelike Killing vector and $\kappa$ denotes the surface gravity.
In addition, the black hole entropy can be obtained using the Wald formula,
\begin{equation}\label{eqS}
S=-2\pi \int_{\mathcal{H}} d^{2}x\,\sqrt{h}\,
\frac{\partial \mathcal{L}}{\partial \mathcal{R}_{abcd}}\,
\epsilon_{ab}\epsilon_{cd},
\end{equation}
where the integral is taken over the bifurcation surface $\mathcal{H}$ of the horizon, $h$ is the determinant of the induced metric on $\mathcal{H}$, and $\epsilon_{ab}$ denotes the binormal to the horizon cross section.
Other thermodynamic potentials associated with the cosmological constant, higher-curvature couplings, and additional fields can be defined through
\begin{equation}\label{eqCh}
\psi_{i}=\int_{\mathcal{H}} \xi_{H}\cdot A_{i},
\end{equation}
where $A_{i}$ represents the corresponding gauge potential, and $\xi_{H}$ is the horizon-generating Killing vector.

Using these definitions, the first law of black hole thermodynamics and the Smarr relation take the form
\begin{align}
dM &= T\,dS + \sum_{i}\psi_{i}\,d\alpha_{i}, \\
M &= 2TS + \sum_{i}\alpha_{i}\,\psi_{i},
\label{Smarr}
\end{align}
where $\alpha_{i}$ denote the associated coupling constants or thermodynamic parameters.

By substituting Eqs.~\eqref{eqTe}, \eqref{eqS}, and \eqref{eqCh} into the Smarr relation~\eqref{Smarr}, the mass of the black hole can be determined. 
\begin{align}
    M(r_{+},\alpha_{1},\alpha_{2},...)=&\dfrac{\sqrt{f_{1}(r_{+},\alpha_{1},\alpha_{2},...)h_{1}(r_{+},\alpha_{1},\alpha_{2},...)}}{2\pi}S(r_{+},\alpha_{1},\alpha_{2},...)\nonumber\\
    &+\sum_{i}\alpha_{i}\,\psi_{i}(r_{+},\alpha_{1},\alpha_{2},...).
\end{align}
Inserting this expression for the mass into the first law of thermodynamics then yields partial differential equations for the near-horizon coefficients $f_{1}$ and $h_{1}$. 

\begin{align}
    &\left.\dfrac{\partial M}{\partial r_{+}}\right\vert_{\alpha_{1},\alpha_{2},...}dr_{+}+\left.\dfrac{\partial M}{\partial \alpha_{1}}\right\vert_{r_{+},\alpha_{2},...}d\alpha_{1}+\left.\dfrac{\partial M}{\partial \alpha_{2}}\right\vert_{r_{+},\alpha_{1},...}d\alpha_{2}+...= T\left.\dfrac{\partial S}{\partial r_{+}}\right\vert_{\alpha_{1},\alpha_{2},...}dr_{+}\nonumber\\
    &+T\left.\dfrac{\partial S}{\partial \alpha_{1}}\right\vert_{r_{+},\alpha_{2},...}d\alpha_{1}+T\left.\dfrac{\partial S}{\partial \alpha_{2}}\right\vert_{r_{+},\alpha_{1},...}d\alpha_{2}+...+\psi_{1}d\alpha_{1}+\psi_{2}d\alpha_{2}+...
\end{align}
yielding
\begin{align}
    &\left.\dfrac{\partial M}{\partial r_{+}}\right\vert_{\alpha_{1},\alpha_{2},...} -T\left.\dfrac{\partial S}{\partial r_{+}}\right\vert_{\alpha_{1},\alpha_{2},...}=0,\\
    &\left.\dfrac{\partial M}{\partial \alpha_{1}}\right\vert_{r_{+},\alpha_{2},...}-T\left.\dfrac{\partial S}{\partial \alpha_{1}}\right\vert_{r_{+},\alpha_{2},...}-\psi_{1}=0,\\
    &\left.\dfrac{\partial M}{\partial \alpha_{2}}\right\vert_{r_{+},\alpha_{1},...}-T\left.\dfrac{\partial S}{\partial \alpha_{2}}\right\vert_{r_{+},\alpha_{1},...}-\psi_{2}=0,\\
    &\qquad\vdots\qquad\qquad\qquad\vdots\qquad\qquad\qquad\vdots\nonumber
\end{align}

as differential equations for $f_{1}(r_{+},\alpha_{1},\alpha_{2},\ldots)$ and $h_{1}(r_{+},\alpha_{1},\alpha_{2},\ldots)$.
To solve these equations, it is necessary to assume a specific relation between $f_{1}$ and $h_{1}$. As shown in Ref.\cite{Bonanno:2019rsq} for Einstein–quadratic gravity, the cases $f_{1}\neq h_{1}$ and $f_{1}=h_{1}$ lead to qualitatively different branches of solutions\cite{Sajadi:2025nkm,Sajadi:2020axg,Sajadi:2025tnn,Sajadi:2022ybs,Ponglertsakul:2026whw,Sajadi:2025svc}. Once the differential equations for $f_{1}$ or $h_{1}$ are solved, the near-horizon expansion of the metric functions is completely determined.
In other papers on this subject, $f_{1}$ and $h_{1}$ are treated as free parameters and determined numerically by fixing them, in such a way by integrating the field equations outward, and tuning them to achieve the desired asymptotic behavior\cite{Lu:2015cqa,Bonanno:2019rsq}. We are now in a position to derive the black hole metric and simultaneously analyze its thermodynamics. In the next section, we examine these results within Einstein gravity.

\subsection{Example}\label{sec2.1}

\begin{figure}[H]
\centering
\includegraphics[width=0.8\columnwidth]{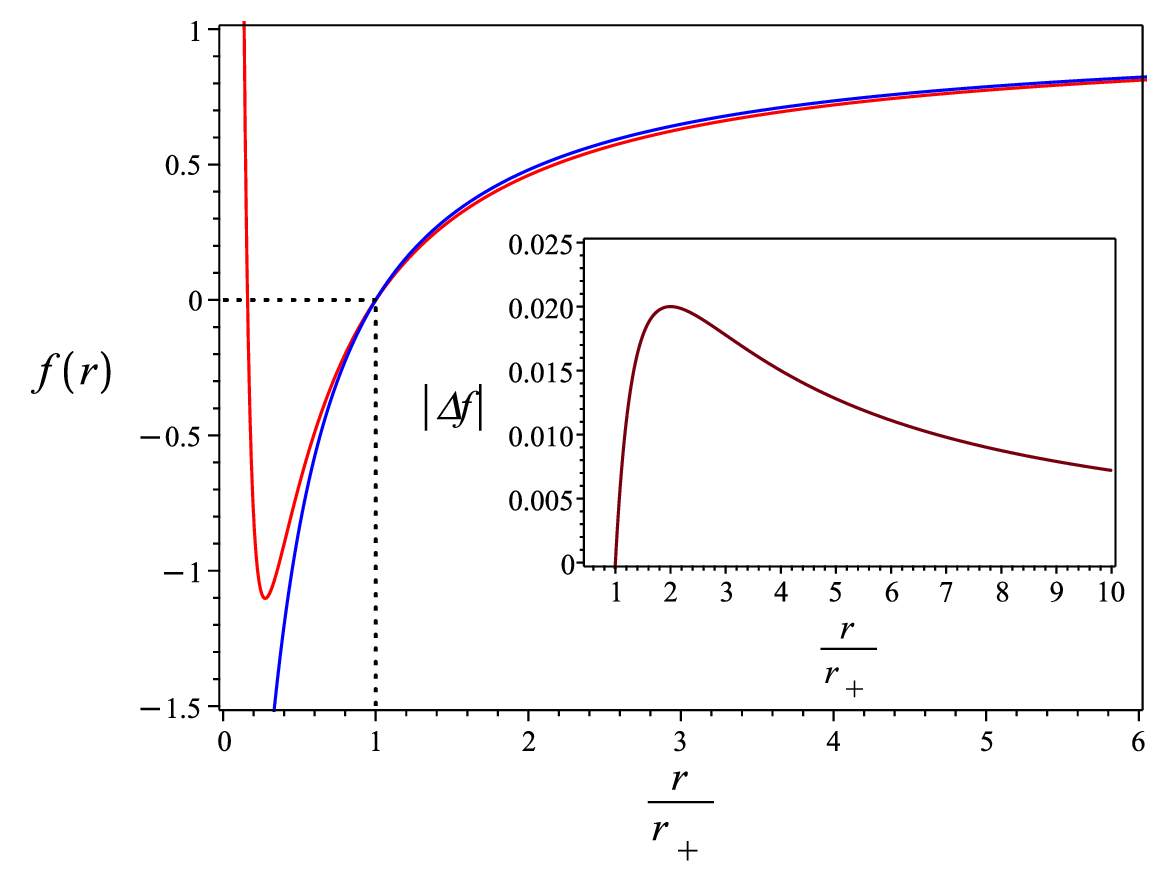}
\caption{Behavior of the metric function. The red curve represents the exact RN solution, while the blue curve shows the result obtained from the continued fraction expansion truncated at four terms, for $q=0.4\; r_{+}$. The inset displays the absolute error between the two curves.}
\label{frplot}
\end{figure}

We start from Einstein–Maxwell with cosmological constant in 4D:
\begin{equation}
    S=\dfrac{1}{16\pi G}\int d^{4}x\sqrt{-g}\left(\mathcal{R}-2\Lambda-\dfrac{1}{4}F_{\mu\nu}F^{\mu\nu}\right)
\end{equation}
the corresponding field equations are given by
\begin{equation}
    G_{\mu\nu}+\Lambda g_{\mu\nu}=F_{\mu\alpha}F_{\nu}^{\alpha}-\dfrac{1}{4}g_{\mu\nu}F_{\alpha\beta}F^{\alpha\beta},\qquad \nabla_{\mu}F^{\mu\nu}=0.
\end{equation}
Assume a static, spherically symmetric spacetime
\begin{equation}
    ds^2=-f(r)dt^2+\dfrac{dr^2}{f(r)}+r^2d\Omega_{2}^2.
\end{equation}
Take a purely electric field
\begin{equation}
    A_{\mu}=\left(\phi(r),0,0,0\right),\qquad F_{tr}=-\phi^{\prime}(r),
\end{equation}
using the Maxwell equation, one can get
\begin{equation}
    F_{tr}=\dfrac{Q}{r^2},
\end{equation}
where $Q$ is the electric charge. The nonzero components Stress–energy tensor are
\begin{equation}
    T^{t}_{t}=T^{r}_{r}=-\dfrac{Q^{2}}{2r^4},\qquad  T^{\theta}_{\theta}=T^{\phi}_{\phi}=\dfrac{Q^{2}}{2r^4}.
\end{equation}
For the metric ansatz, the $rr-$component of Einstein equation is given by
\begin{equation}
    \dfrac{f^{\prime}r+f-1}{r^2}+\Lambda=-\dfrac{Q^2}{2r^4}
\end{equation}
By solving the above first-order equation, one can obtain the RN–AdS black hole solution. Here, we instead employ the continued fraction expansion described in the previous section to construct the black hole solution.
Therefore, we consider the following asymptotic metric 
\begin{equation}
    f(r)=\sum_{i=0}\dfrac{F_{i}}{r^{i}}=1+\dfrac{F_{1}}{r}+\dfrac{Q^2}{r^2}-\dfrac{\Lambda}{3}r^2
\end{equation}
and near the horizon 
\begin{align}
    f(r)=&\sum_{i=1}f_{i}(r-r_{+})^i=f_{1}(r-r_{+})+\dfrac{(Q^2-r_{+}^2)}{r_{+}^4}(r-r_{+})^2\nonumber\\
    &-\dfrac{(4 Q^2-f_{1} r_{+}^3-2 r_{+}^2)}{3r_{+}^5}(r-r_{+})^3+\mathcal{O}(r-r_{+})^4.
\end{align}
The coefficients appearing in the continued fraction expansion are obtained as follows ($\Lambda=0$):
\begin{align}
\epsilon=&-\dfrac{F_{1}}{r_+}-1,  \qquad a_{0}=\dfrac{Q^{2}}{r_{+}^{2}},\qquad  a_1=-1-a_{0}+2\epsilon+r_{+}f_1, \nonumber\\
a_{2}=&-\dfrac{1}{ a_1} \left[4a_1-5\epsilon+1+3 a_{0}+ f_{2}r_{+}^2\right], \nonumber\\
a_{3}=&-\dfrac{1}{{a_{1}}{a_{2}}}[-{f_{3}}{{r_{+}}}^{3}+{a_{1}}{{a_{2}}}^{2}+5{a_{1}}{a_{2}}+6{a_{0}}+10{a_{1}}-9\epsilon+1].
\end{align}
In the above coefficients, all quantities are determined except $f_{1}$ and $F_{1}=-2M$.
To determine them, we use thermodynamic considerations, as in the previous section.
The thermodynamical quantities are given as
\begin{equation}
    T=\dfrac{f_{1}}{4\pi},\quad S=\pi r_{+}^2,\quad P=-\dfrac{\Lambda}{8\pi},\quad V=\dfrac{4}{3}\pi r_{+}^3,\quad \phi=\dfrac{Q}{r_{+}}.
\end{equation}
Therefore, the Smarr formula yields
\begin{equation}
    M=2TS-2PV+Q\phi=\dfrac{r_{+}^2}{2} f_{1}(r_{+},\Lambda,Q)+\dfrac{\Lambda}{3}r_{+}^3+\dfrac{Q^2}{r_{+}}
\end{equation}
and substituting them into the first law of thermodynamics,
\begin{align}
dM=TdS+VdP+\phi dQ=
\begin{cases}  
&\frac{1}{2}\, r_{+}^2 \frac{\partial f_1(r_{+},\Lambda,Q)}{\partial r_{+}}
+ \frac{1}{2}\, r_{+}\, f_1(r_{+},\Lambda,Q)
+ \Lambda r_{+}^2
- \frac{Q^2}{r_{+}^2}=0,
\\[6pt]
&\frac{1}{2}\, r_{+}^2 \frac{\partial f_1(r_{+},\Lambda,Q)}{\partial \Lambda}
+ \frac{1}{2}\, r_{+}^3=0,
\\[6pt]
&\frac{1}{2}\, r_{+}^2 \frac{\partial f_1(r_{+},\Lambda,Q)}{\partial Q}
+ \frac{Q}{r_{+}}=0.\nonumber
\end{cases}
\end{align}
One can obtain the following by solving the above differential equations simultaneously:
\begin{equation}
f_1(r_{+},\Lambda,Q)
=
- r_{+}\,\Lambda
- \frac{Q^2}{r_{+}^3}
+ \frac{C_1}{r_{+}}.
\end{equation}
So, the temperature becomes
\begin{equation}
    T=\frac{f_1(r_{+},\Lambda,Q)}{4\pi}=\dfrac{1}{4\pi}\left[- r_{+}\,\Lambda
- \frac{Q^2}{r_{+}^3}
+ \frac{C_1}{r_{+}}\right].
\end{equation}
For $C_1=1$, this exactly reproduces the temperature of the RN–AdS black hole. Therefore,
\begin{align}
f_1(r_{+},\Lambda,Q)
=&
- r_{+}\,\Lambda
- \frac{Q^2}{r_{+}^3}
+ \frac{1}{r_{+}},\nonumber\\
    F_{1}=-2M=&-r_+
\left[
1
+
\frac{Q^2}{r_+^2}
-
\frac{\Lambda r_+^2}{3}
\right].
\end{align}
Finally, using continued fraction expansion, the metric can be readily obtained.
In figure \ref{frplot}, we have shown the metric function of the RN black hole obtained from the exact solution (red curve) together with the result reconstructed using the continued fraction expansion truncated at four terms (blue curve). The inset plots the deviation between the exact RN metric function and the continued-fraction approximation.
One observes excellent agreement between the two curves for the outside of black hole ($\Delta f\le 10^{-2}$). In particular, the continued fraction approximation accurately reproduces both the near-horizon behavior and the asymptotic of the exact solution. This demonstrates that even a low-order truncation of the continued fraction expansion is sufficient to capture the essential features of the RN geometry with high accuracy.

\section{Conclusions}\label{con}

{Since higher-curvature corrections generally lead to highly nonlinear and complicated field equations, obtaining analytical black hole solutions directly from the equations of motion remains challenging. In this work, we demonstrated that combining black hole thermodynamics with the continued fraction expansion provides an efficient framework for constructing approximate analytical black hole solutions in higher-curvature gravity theories. Focusing on Einstein gravity, we showed that the first law of thermodynamics and the Smarr relation are sufficient to determine black hole solutions and their thermodynamic properties without directly solving the field equations. The resulting thermodynamic quantities and black hole solutions are in agreement with those obtained from the exact equations of motion, confirming the validity and consistency of the approach. Since the continued fraction expansion basically acts as a parametrization of the metric functions and has only limited dependence on the underlying gravitational theory, its applicability is not restricted to higher-curvature gravity. This suggests that the framework can be extended to non-curvature-based modified gravity theories, such as $f(Q)$ and $f(T)$ gravity, opening a new window for constructing analytical black hole solutions in a broader class of gravitational theories. A detailed investigation of such extensions is left for future work.}

\section*{Acknowledgments}
This research has received funding support from the NSRF via the Program Management Unit
for Human Resource and Institutional Development, Research and Innovation grant number
B13F680083.


\begin{thebibliography}{99}

\bibitem{Lu:2015cqa}
H.~Lu, A.~Perkins, C.~N.~Pope and K.~S.~Stelle,
Phys. Rev. Lett. \textbf{114}, no.17, 171601 (2015)


\bibitem{Myers:1998gt}
R.~C.~Myers,
doi:10.1007/978-94-017-0934-7{\_}8

\bibitem{Capozziello:2022ygp}
S.~Capozziello, R.~D'Agostino, A.~Lapponi and O.~Luongo,
Eur. Phys. J. C \textbf{83}, no.2, 175 (2023)

\bibitem{Luongo:2015zaa}
O.~Luongo and H.~Quevedo,
Found. Phys. \textbf{48}, no.1, 17-26 (2018)

\bibitem{Luongo:2014qoa}
O.~Luongo and H.~Quevedo,
Phys. Rev. D \textbf{90}, no.8, 084032 (2014)

\bibitem{Luongo:2010we}
O.~Luongo and H.~Quevedo,
doi:10.1142/9789814374552{\_}0122

\bibitem{Luongo:2023xaw}
O.~Luongo, H.~Quevedo and S.~N.~Sajadi,
Gen. Rel. Grav. \textbf{56}, no.2, 17 (2024)

\bibitem{Sajadi:2025prp}
S.~N.~Sajadi, S.~Ponglertsakul and O.~Luongo,
Phys. Dark Univ. \textbf{48}, 101938 (2025)

\bibitem{Bonanno:2019rsq}
A.~Bonanno and S.~Silveravalle,
Phys. Rev. D \textbf{99}, no.10, 101501 (2019)


\bibitem{Rezzolla:2014mua}
  L.~Rezzolla and A.~Zhidenko,
  Phys.\ Rev.\ D {\bf 90}, no. 8, 084009 (2014)
 
 


\bibitem{Kokkotas:2017zwt}  
 K.~Kokkotas, R.~A.~Konoplya and A.~Zhidenko,  
 Phys.\ Rev.\ D {\bf 96}, no. 6, 064007 (2017)   
 
   
 
\bibitem{Konoplya:2019ppy} 
 R.~A.~Konoplya and A.~F.~Zinhailo,  
 Phys.\ Rev.\ D {\bf 99}, no. 10, 104060 (2019)  


  
\bibitem{Sajadi:2025nkm}
S.~N.~Sajadi and S.~Ponglertsakul,
Annals Phys. \textbf{477}, 170007 (2025)


\bibitem{Sajadi:2020axg}
S.~N.~Sajadi, R.~B.~Mann, N.~Riazi and S.~Fakhry,
doi:10.1103/PhysRevD.102.124026


\bibitem{Sajadi:2025tnn}
S.~N.~Sajadi and S.~Ponglertsakul,
Eur. Phys. J. Plus \textbf{140}, no.6, 565 (2025)


\bibitem{Sajadi:2022ybs}
S.~N.~Sajadi and S.~H.~Hendi,
Eur. Phys. J. C \textbf{82}, no.8, 675 (2022)

\bibitem{Ponglertsakul:2026whw}
S.~Ponglertsakul and S.~N.~Sajadi,
Front. Astron. Space Sci. \textbf{13}, 1754814 (2026)

\bibitem{Sajadi:2025svc}
S.~N.~Sajadi, S.~Ponglertsakul and R.~B.~Mann,
Phys. Rev. D \textbf{113}, no.6, 064043 (2026)

\bibitem{Sajadi:2025kah}
S.~N.~Sajadi, S.~Ponglertsakul and D.~J.~Gogoi,
Eur. Phys. J. C \textbf{85}, no.9, 943 (2025)

\bibitem{Zinhailo:2018ska}   
A.~F.~Zinhailo,  
  Eur.\ Phys.\ J.\ C {\bf 78}, no. 12, 992 (2018) 
  

\bibitem{Cai:2015emx}
Y.~F.~Cai, S.~Capozziello, M.~De Laurentis and E.~N.~Saridakis,
Rept. Prog. Phys. \textbf{79}, no.10, 106901 (2016)


\bibitem{Heisenberg:2023lru}
L.~Heisenberg,
Phys. Rept. \textbf{1066}, 1-78 (2024)


\bibitem{Konoplya:2022kld}
R.~A.~Konoplya and A.~Zhidenko,
JCAP \textbf{11}, 028 (2022)
doi:10.1088/1475-7516/2022/11/028
[arXiv:2210.04314 [gr-qc]].


\end{thebibliography}
\end{document}